\documentclass[pra,twocolumn,floatfix,a4paper,superscriptaddress]{revtex4}

\usepackage{bm,color,amsmath,txfonts}

\usepackage{graphicx}
\usepackage{subfigure}
\usepackage{verbatim}
\usepackage{dcolumn}
\usepackage{bm}
\usepackage{epsf}
\usepackage{xcolor}
\usepackage{hyperref}
\usepackage{hhline}
\usepackage{float}
\usepackage{enumerate}
\usepackage{bbm}
\usepackage{lipsum}
\definecolor{mygreen}{rgb}{0.01, 0.31, 0.59}
\definecolor{myblue}{rgb}{0.01, 0.31, 0.59}
\hypersetup{
 colorlinks=true,   
 linkcolor=blue,  
 citecolor=blue, 
 urlcolor =blue    
}

\usepackage{hyperref}
\newcommand{\da}{^\dagger}
\newcommand{\non}{\nonumber}
\newcommand{\cbr}[1]{\left( #1 \right)}

\newcommand{\ee}{{\rm e}}
\newcommand{\mean}[1]{\langle #1 \rangle}

\begin{document}

\title{Magnon squeezing enhanced ground-state cooling in cavity magnomechanics}

\author{M. Asjad} \thanks{asjad\_qau@yahoo.com}
\affiliation{Department of Applied Mathematics and Sciences, Khalifa University, Abu Dhabi 127788, United Arab Emirates}
\author{Jie Li} \thanks{jieli007@zju.edu.cn}
\affiliation{Interdisciplinary Center of Quantum Information, State Key Laboratory of Modern Optical Instrumentation, and Zhejiang Province Key Laboratory of Quantum Technology and Device, Department of Physics, Zhejiang University, Hangzhou 310027, China}
\author{Shi-Yao Zhu}
\affiliation{Interdisciplinary Center of Quantum Information, State Key Laboratory of Modern Optical Instrumentation, and Zhejiang Province Key Laboratory of Quantum Technology and Device, Department of Physics, Zhejiang University, Hangzhou 310027, China}
\author{J. Q. You}
\affiliation{Interdisciplinary Center of Quantum Information, State Key Laboratory of Modern Optical Instrumentation, and Zhejiang Province Key Laboratory of Quantum Technology and Device, Department of Physics, Zhejiang University, Hangzhou 310027, China}

\begin{abstract}
Cavity magnomechanics has recently become a new platform for studying macroscopic quantum phenomena. The magnetostriction induced vibration mode of a large-size ferromagnet or ferrimagnet reaching its ground state represents a genuine macroscopic quantum state. Here we study the ground-state cooling of the mechanical vibration mode in a cavity magnomechanical system, and focus on the role of magnon squeezing in improving the cooling efficiency. {The magnon squeezing is obtained by exploiting the magnon self-Kerr nonlinearity}. We find that the magnon squeezing can significantly and even completely suppress the magnomechanical Stokes scattering. It thus becomes particularly useful in realizing ground-state cooling in the unresolved-sideband regime, where the conventional sideband cooling protocols become inefficient. We also find that the coupling to the microwave cavity plays only an adverse effect in mechanical cooling. This makes essentially the two-mode magnomechanical system (without involving the microwave cavity) a preferred system for cooling the mechanical motion, in which the magnon mode is established by a uniform bias magnetic field and a microwave drive field. \\

\end{abstract}

\date{\today}
\maketitle

\section*{1. Introduction}

Cooling a massive mechanical oscillator into its quantum ground state finds many applications not only in fundamental researches related to macroscopic quantum phenomena (e.g., the quantum-to-classical transition and the test of unconventional decoherence theories), but also in high-sensitivity measurements and quantum information processing~\cite{omRMP,DV09,Bassi,ORI,Liu}. Cavity optomechanics, studying the interaction between light and mechanical motion by radiation pressure, provides an ideal platform to achieve this goal. In the past decade, cavity optomechanics has witnessed successful experimental demonstrations of the quantum ground state of mechanical motion~\cite{c1,c2,c4,c5,c6,c7,c8}, by exploiting different cooling mechanisms, such as measurement-based feedback cooling~\cite{fb1,fb2}, sideband cooling~\cite{sb1,sb2,sb3}, and cavity-enhanced coherent scattering~\cite{cs1,cs2}.
 
In recent years, the analogous cavity magnomechanical system has received considerable attention~\cite{Naka19,Yuan}. It has been shown that cavity magnomechanics (CMM) has the potential to prepare macroscopic quantum states~\cite{Jie18,Jie19a,Jie19b,Tan,CLi,Jie21,Wang,Wu,CLi2} and nonclassical states of microwave fields~\cite{Jie20,NSR}, and it has also promising applications in quantum information processing and quantum technologies~\cite{Tang,Xiong,Davis,Qi1,Sarma,HJ,Davis2}. In the CMM, a magnon mode (spin wave) couples to a deformation vibration mode of a ferromagnet (or ferrimagnet) via the magnetostrictive force, and to a microwave cavity mode via the magnetic dipole interaction. The magnetostrictive interaction is a radiation pressure-like dispersive interaction for a {large-size} ferromagnet, where the frequency of the mechanical vibration mode is much lower than the magnon frequency~\cite{Tang,Jie22}. The similarity between the magnomechanical and optomechanical interactions allows one to apply optomechanical cooling mechanisms to the CMM. Using the wisdom of sideband cooling~\cite{sb1,sb2,sb3}, the vibration mode can be efficiently cooled (close) to its quantum ground state by driving the magnon mode with a red-detuned microwave field~\cite{Jie18,Jie19a,CLi}, which is a precondition for observing entangled and squeezed-vacuum states in the CMM~\cite{Jie18,Jie19a,Jie19b,Jie20}. Directly driving magnons by a microwave field, e.g., via a loop antenna~\cite{YP} (compared to indirectly driving by the microwave cavity field~\cite{Tang}), yields a higher pump efficiency~\cite{NSR} and a more direct mapping between magnomechanics and optomechanics. In {all} of the above theoretical and experimental studies~\cite{Jie18,Jie19a,Jie19b,Tan,CLi,Jie21,Wang,Wu,CLi2,Jie20,NSR,Tang,Xiong,Davis,Qi1,Sarma,HJ,Davis2}, the resolved sidebands are assumed, which is well satisfied for ferrimagnetic yttrium iron garnet (YIG), because it has a very low magnon damping rate. However, for some ferromagnetic materials, e.g., CoFeB~\cite{prb}, though the magnomechanical coupling strength can be much stronger, the magnon damping is much larger, entering the unresolved-sideband regime, and thus the conventional sideband-cooling protocols become inefficient. Therefore, providing efficient cooling schemes for magnomechanical systems with {unresolved sidebands} are highly needed.

To address this issue, here we provide a mechanism for cooling the vibration mode in a CMM system by utilizing the squeezing of the magnon mode. We focus on the role of magnon squeezing in improving the cooling efficiency. {The magnon squeezing can be achieved by utilizing the magnon self-Kerr nonlinearity}. We find that the magnon squeezing can significantly and even fully suppress the magnomechanical Stokes scattering, giving rise to a remarkably enhanced cooling efficiency. This makes the ground-state cooling possible in the unresolved-sideband regime. We also find an adverse role of the coupling to the microwave cavity in mechanical cooling: A larger magnon-cavity coupling yields a larger effective mean phonon number. This means, essentially, the more compact two-mode magnomechanical system~\cite{Jie22} (without a microwave cavity) is preferred, compared to the three-mode CMM system, for the purpose of mechanical cooling.


\begin{figure}[t]
\hskip0.3cm\includegraphics[width=0.85\columnwidth]{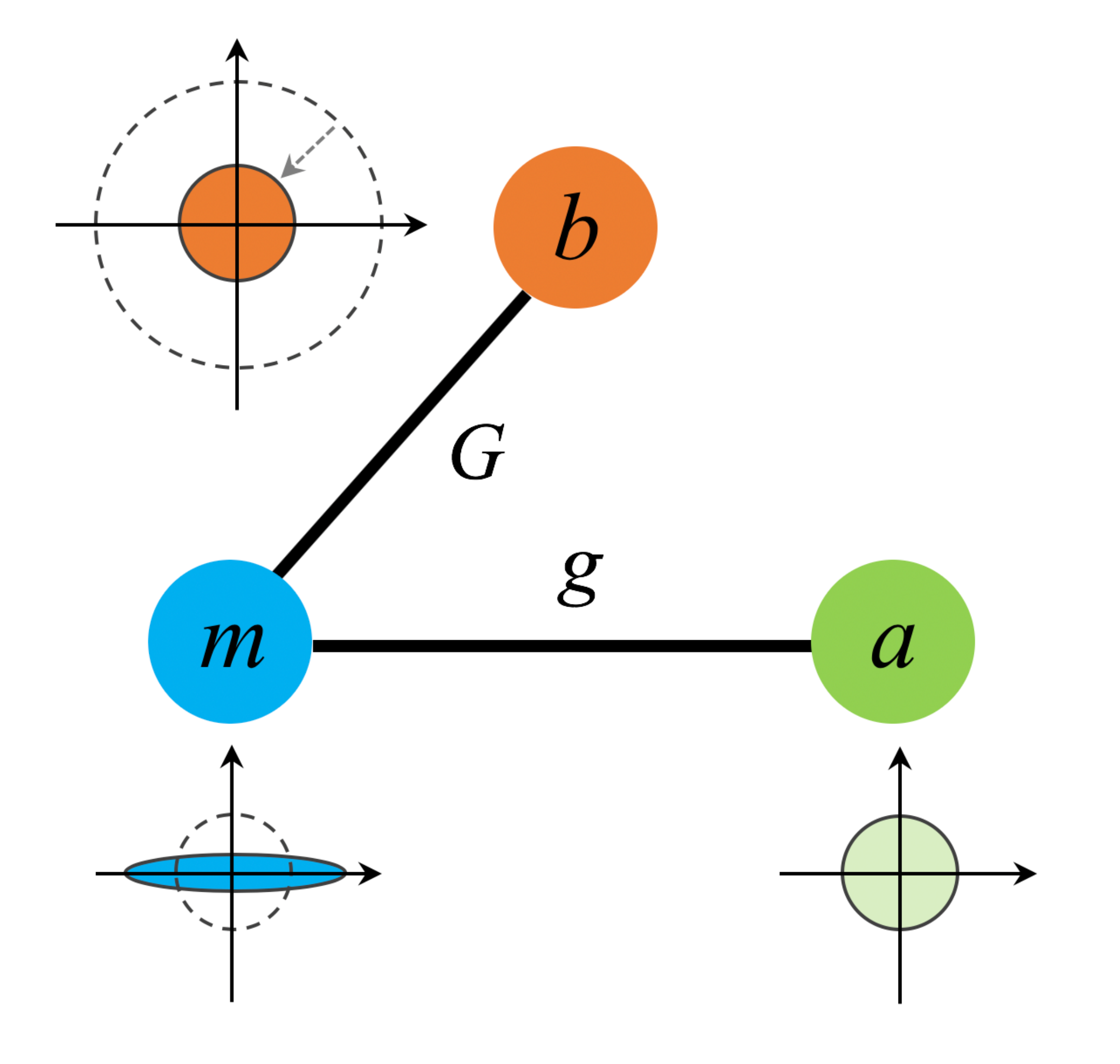}\\
\caption{The cavity-magnomechanical system consisting of a magnon mode ($m$), a mechanical vibration mode ($b$), and a microwave cavity mode ($a$). The colored circles and ellipse in phase space represent the noise/fluctuation of the quadratures of the three modes. The squeezing of the magnon mode (sketched by the blue ellipse) is proved to be useful in improving the cooling of the mechanical mode (denoted by the circle with reduced radius) and in achieving its quantum ground state in the unresolved-sideband limit. }
\label{fig1}
\end{figure}

\section*{2. The model}\label{model}

The three-mode CMM system consists of a magnon mode, a microwave cavity mode, and a mechanical vibration mode, as shown in Fig.~\ref{fig1}. The magnons are the quanta of a spin wave in a magnetically ordered state. A magnon mode can be the uniform-precession Kittel mode in a ferrimagnetic YIG sphere~\cite{Tang,Davis2}. The magnon mode simultaneously couples to a microwave cavity mode via the magnetic dipole interaction and to a deformation vibration mode via the magnetostrictive interaction. Owing to a large size of the ferrimagnet (typically in the range from $10^2$ $\mu$m to 1 mm), the frequency of the vibration mode is much lower than the magnon frequency, which yields a dominant dispersive coupling between magnons and phonons~\cite{Jie22}. The Hamiltonian of the CMM system, in the frame rotating at the magnon drive frequency $\omega_{0}$, reads

 \begin{eqnarray}
  H/\hbar \!\!&{=}&\!\!  \Delta_a a^\dagger a + \Delta_m m^\dagger m + \omega_b b^\dagger b + g (a^\dagger m+a m^\dagger) \non\\
  &{+}& \!\! G_0 m^\dagger m (b+b^\dagger) +{\xi m^\dagger m m^\dagger m  + i ( E m^\dagger - E^* m )},
 \end{eqnarray}
where $a$, $m$, and $b$ ($a^\dag$, $m^\dag$, and $b^\dag$) are the annihilation (creation) operators of the cavity, magnon, and mechanical modes, respectively, $\omega_{k}$ ($k=a,m,b$) is the resonance frequency of the corresponding mode, and $\Delta_{a(m)}=\omega_{a(m)}-\omega_0$. $g$ denotes the magnon-cavity coupling strength, and $G_{0}$ is the bare magnomechanical coupling rate, which can be significantly improved by directly driving the magnon mode with a microwave field with the coupling strength $E \,{=}\, |E| e^{i \theta}$, and $\theta$ being the phase of the drive field. The magnon self-Kerr term, $\hbar \xi m^\dagger m m^\dagger m$ ($\xi$ is the self-Kerr coefficient), is the novel part with respect to the conventional CMM systems, which is responsible for magnon squeezing. It is known that self-Kerr systems can produce squeezing~\cite{Walls,Rebic}. The magnon self-Kerr nonlinearity can be achieved by coupling the magnon mode to a superconducting qubit~\cite{Quirion1}. This coupling yields an effective magnon self-Kerr nonlinearity~\cite{Naka19}. The strength of the squeezing increases with the magnon self-Kerr coefficient $\xi$, which can be tuned by adjusting the effective coupling between the magnons and the qubit. This can be easily realized by changing their respective couplings to the microwave cavity. The magnon self-Kerr nonlinearity can also be obtained by using the magnetocrystalline anisotropy in YIG~\cite{prb16}, and the Kerr coefficient can be tuned by adjusting the angle between the bias magnetic field and the crystal axis of YIG~\cite{NSR}.

The full dynamics of the system can be described by the Heisenberg equations of motion and by adding the corresponding damping and noise terms, which lead to the following quantum Langevin equations (QLEs):
\begin{eqnarray}
\dot{a}&=&-(\gamma_a+i\Delta_a)a-i g m +\sqrt{2\gamma_a} a_{in}, \non \\
\dot{b}&=&-(\gamma_b+i\omega_b)b - i G_0 m\da m+\sqrt{2\gamma_b} b_{in},\non\\
\dot{m}&=&-\cbr{\gamma_m+i\Delta_m} m -i g a - i G_0 (b+b\da)m  - {2i \xi  m\da m m} \non \\ 
&+&\! \! E + \sqrt{2\gamma_m} m_{in},\label{HL}
\end{eqnarray}
where $\gamma_k$ ($k=a,b,m$) is the damping rate of the corresponding mode, and $k_{in}(t)$ denotes the input noise, which is zero-mean and delta-correlated: $\mean{k_{in}(t) k\da_{in}(t')} =(n_{k}+1)\delta(t-t')$ and $\mean{k\da_{in}(t) k_{in}(t')} =n_{k}\delta(t-t')$, with  $n_{k}=\big( \ee^{\hbar \omega_{k}/k_B T}-1\big)^{-1}$ being the mean thermal occupation number of the corresponding mode. For the mechanical noise, we take a Markovian approximation leading to a delta noise correlation, which is valid for a large mechanical quality factor $Q_b=\omega_b/\gamma_b \gg 1$~\cite{Jie18}.

\begin{figure*}[t!] 
\includegraphics[width=0.99\textwidth]{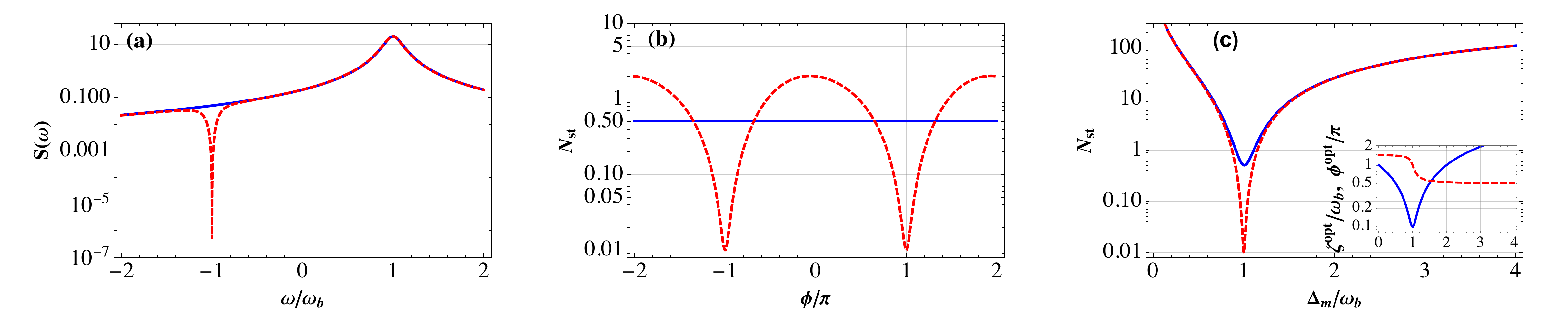}\\
\includegraphics[width=0.99\textwidth]{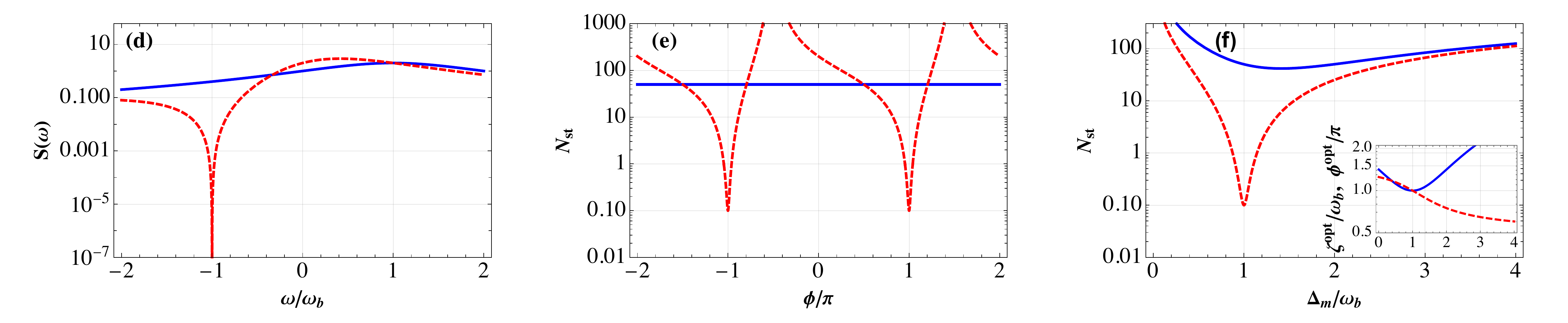}\\
\includegraphics[width=0.99\textwidth]{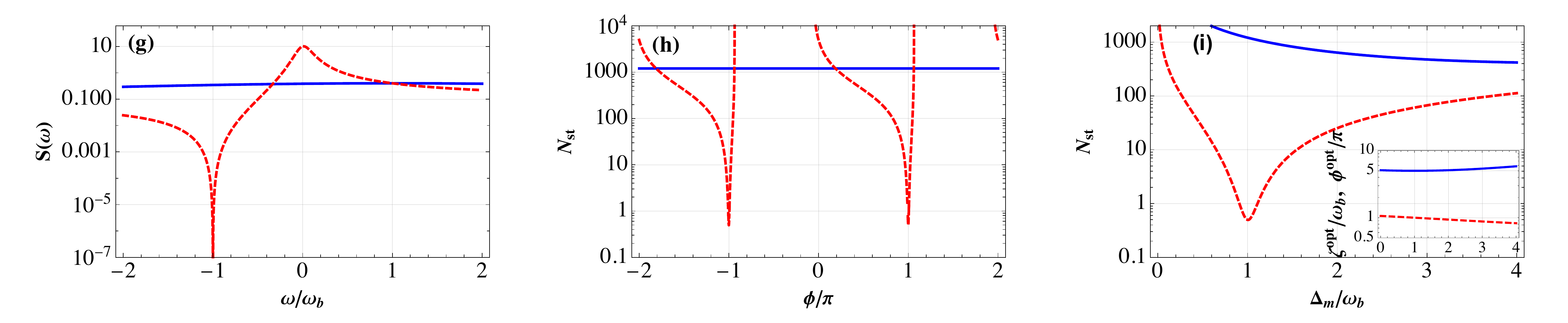}
\caption{First column (a), (d) and (g): Power spectrum of the magnon mode $S(\omega)$ versus the normalized frequency $\omega/\omega_b$, with $\Delta_m = \omega_b$ and optimal values of $|\zeta|^{\rm opt}$ and $\phi^{\rm opt}$. Second column (b), (e) and (h): Steady-state effective mean phonon number $N_{st}$ versus the phase $\phi/\pi$, with $\Delta_m = \omega_b$ and optimal $|\zeta|^{\rm opt}$. Third column (c), (f) and (i): $N_{st}$ versus the normalized detuning $\Delta_m/\omega_b$ with optimal squeezing parameters $\zeta^{\rm opt}$ and $\phi^{\rm opt}$ (Inset: $\zeta^{\rm opt}$ -- blue solid line; $\phi^{\rm opt}$ -- red dashed line). Three rows from upper to lower are, respectively, for three values of magnon damping rate: $\gamma_m= 0.1\omega_b$ (resolved sidebands), $\gamma_m= \omega_b$, and $\gamma_m= 5 \omega_b$ (unresolved sidebands). The blue solid (red dashed) curves in the main panels are plotted for the case without squeezing, i.e., $|\zeta| = 0$ (optimal squeezing parameters $\zeta^{\rm opt}$ and $\phi^{\rm opt}$). The other parameters are $n_{b} = 100$, $n_{a(m)}= 0$, $G = 0.1 \omega_b$, $\gamma_a= \omega_b$, $g=0$, and $\gamma_b = 10^{-5} \omega_b$. }
\label{fig2}
\end{figure*}

Under the strong pump for the magnon mode, the QLEs (\ref{HL}) can be linearized by writing each operator $O$ ($O=a,b,m$) as the sum of its steady-state mean value and the corresponding small fluctuation operator with zero mean value, $O\rightarrow \delta O+ O_s$. The QLEs (\ref{HL}) are then separated into two sets of equations for classical averages and for quantum fluctuations, respectively.  The steady-state averages can be obtained by solving the following equations
\begin{eqnarray}
a_s&=& -i g m_s/(\gamma_a+i\Delta_a), \non \\ 
b_s&=&-i G_0 |m_s|^2/(\gamma_b+i\omega_b), \non \\ 
m_s &=& {(E- i g a_s) /(\gamma_m+i\tilde{\Delta}_m)}, 
\end{eqnarray}
where $\tilde{\Delta}_m = \Delta_m + 2 G_0 {\rm Re}\,b_s+{2 \xi |m_s|^2}$ is the effective detuning by including the magnetostriction and {the magnon self-Kerr} induced frequency shifts. Typically, these frequency shifts are small~\cite{Tang,Davis2,Naka19}, i.e., $|\tilde{\Delta}_m - \Delta_m| \ll \Delta_m \simeq \omega_b$ (the optimal detuning for mechanical cooling $\Delta_m \simeq \omega_b$ will be shown later). Therefore, hereafter we can assume $\tilde{\Delta}_m \simeq \Delta_m$.  The explicit expression of the steady-state average $m_s $ is given by
\begin{equation}
m_s =  \frac{ E (\gamma_a + i \Delta_a)}{ g^2  +  ( \gamma_m + i \Delta_m) ( \gamma_a + i \Delta_a) },
\end{equation}
which is generally complex, and its phase can be adjusted by varying the phase of the drive field.

The corresponding linearized QLEs describing the dynamics of the quantum fluctuations are given by (for simplicity, we use $O$ to denote fluctuation operators $\delta O$)
\begin{eqnarray}
\dot{a}&=&-(\gamma_a+i\Delta_a) a- i g  m+\sqrt{2\gamma_a}a_{in},\non\\
\dot{b}&=&-(\gamma_b+i\omega_b) b - { i  (G^* m+ G m\da) }+\sqrt{2\gamma_b}b_{in},\non\\
\dot{m}&=&-(\gamma_m+i\Delta_m) m - i g  a -i G (b+b^\dagger) + { \zeta m\da } + \sqrt{2\gamma_m} m_{in},\non\\
\label{bm}
\end{eqnarray}
where $G=G_0 m_s$ is the effective magnomechanical coupling strength, and $\zeta \equiv -2i \xi m^2_s$ represents the effective squeezing parameter of the magnon mode, with the amplitude $|\zeta|$ and the phase $\phi=\tan^{-1}[ {\rm Im} \, \zeta / {\rm Re} \, \zeta ] $. 

\section*{3. Magnon-squeezing enhanced mechanical cooling}

The radiation pressure-like dispersive coupling to magnons provides a mean to cool the mechanical motion by properly driving the magnon mode~\cite{Jie19a,Jie21,CLi}. The magnon mode then acts as an effective reservoir which competes with the mechanical natural thermal reservoir to determine the stationary energy of the mechanical oscillator. The effective stationary mean phonon number, in the weak coupling limit $|G| \ll \omega_b$, is given by~\cite{sb1,sb2,sb3}
\begin{equation}\label{Nst}
N_{st}=\dfrac{\gamma_b n_b+A_+}{\gamma_b+\Gamma_b},
\end{equation}
where $\Gamma_b= A_{-}-A_+$ is the net cooling rate at which mechanical excitations are dissipated into an effective magnon-controlled reservoir, and $A_{\pm}$ define the rates at which the driving microwave photons are scattered by the mechanical motion simultaneously with the absorption (the anti-Stokes scattering with rate $A_-$) or emission (the Stokes scattering with rate $A_+$) of the vibrational phonons. For a red-detuned drive, $\Delta_m>0$, one has  $A_{-} >A_+$, and thus $\Gamma_b= A_{-}-A_+>0$. {The system enters a regime where the anti-Stokes process by absorbing phonons outperforms the Stokes process by emitting phonons. Therefore, the mechanical mode is cooled with the final phonon occupancy $N_{st} < n_b$ when $n_b \gg 1$.} This is termed as the sideband cooling, which works optimally at the detuning $\Delta_m = \omega_b$ and in the resolved-sideband limit $\omega_b \gg \kappa_m$~\cite{sb1,sb2,sb3}.

Equation~\eqref{Nst} implies that the effective mean phonon number $N_{st}$ can be significantly reduced by suppressing the Stokes scattering rate $A_+ \,{\to}\,\, 0$.  Enlightened by the fact that squeezed light can be used to suppress the optomechanical Stokes scattering~\cite{c4,Asjad,Asjad2,YC,Lau}, we apply the idea to the magnomechanical system and expect that the magnomechanical Stokes scattering would be suppressed by utilizing the squeezing of the magnon mode. The Stokes and anti-Stokes scattering rates are given by $A_{\pm}= |G|^2 S(\mp\omega_b)$, where 
\begin{equation}
S(\omega) =\int_{-\infty}^\infty \mean{ X(t) \, X(0)} e^{i\omega t} dt
\end{equation} 
is the spectrum of fluctuations of the magnon field quadrature $X=(m+m^\dagger)$, which couples to the mechanics at $\omega= \mp \omega_b$. By taking the Fourier transform of Eq.~\eqref{bm} and solving it in the frequency domain, we obtain the explicit expression for the power spectrum
\begin{eqnarray}
S(\omega)&=& 2\dfrac{\left(\gamma_m+ g^2 \gamma_a |\chi_a(\omega)|^2\right) \left||\zeta| e^{i\phi}+\mathcal{A}(-\omega) \right|^2}{
\left|\mathcal{A}(\omega) \mathcal{A}^*(-\omega) -|\zeta|^2 \right|^2},
\end{eqnarray}
where $\mathcal{A}(\omega)= \chi^{-1}_m(\omega)  + g^2 \chi_{a}(\omega)$, and the natural susceptibility of the cavity (magnon) mode is given by $\chi_{a\,(m)}(\omega)=\big[ \gamma_{a\,(m)} + i(\Delta_{a\,(m)}-\omega)\big]^{-1} $.

A key observation is that the Stokes scattering, accounting for the heating of the mechanical mode, can be {fully} suppressed, i.e, 
\begin{equation}
A_{+}(-\omega_b)|_{\Delta_{a}=\omega_b}=0,  
\end{equation}
under the conditions 
\begin{eqnarray}
|\zeta| \!\!&{=}&\!\! |\zeta|^{\rm opt}=\sqrt{\gamma'^{2}_m +(\Delta_m-\omega_b)^2}, \,\,\,\,\,\,\,\,  {\rm and} \non  \\  
\phi \!\!&{=}&\!\! \phi^{\rm opt}=-i\ln\left[-\big\{ \gamma'_m-i (\Delta_m-\omega_b) \big\} / |\zeta|^{\rm opt}\right],  \label{opt2}
\end{eqnarray}
where $\gamma'_m=\gamma_m+g^2/\gamma_a$. This is clearly shown in Fig.~\ref{fig2}(a), \ref{fig2}(d) and \ref{fig2}(g) for three values of the magnon damping rate from the resolved-sideband regime ($\gamma_m \ll \omega_b$) to the unresolved-sideband regime ($\gamma_m \gg \omega_b$), embodied by the vanishing power spectrum at $\omega=-\omega_b$, i.e., $S(-\omega_b) \to 0$. The significantly suppressed Stokes scattering leads to a remarkably enhanced net cooling rate $\Gamma_b$ and thus a significantly reduced mean phonon number $N_{st}$ at the optimal detuning $\Delta_m = \omega_b$, as displayed in Fig.~\ref{fig2}(c), \ref{fig2}(f) and \ref{fig2}(i). The insets of Fig.~\ref{fig2}(c), \ref{fig2}(f) and \ref{fig2}(i) show the corresponding optimal values of the magnon squeezing parameters $|\zeta|$ and $\phi$. For the case of no magnon-cavity coupling $g=0$ and $\Delta_m = \omega_b$, one gets from Eq.~\eqref{opt2} that $ |\zeta|^{\rm opt} =\gamma_m$ and $\phi^{\rm opt} = (2n+1) \pi$, $n\in Z $. The optimal phase is more clearly shown in Fig.~\ref{fig2}(b), \ref{fig2}(e) and \ref{fig2}(h). We take the initial mean thermal occupation number $n_b=100$, corresponding to the bath temperature about 48 mK for a $\sim$10 MHz mechanical mode~\cite{Tang,Davis2}. At this temperature, the thermal occupancy for a $\sim$10 GHz magnon (cavity) mode is $n_{m(a)} \simeq 0$. It is worth noting that, even in the unresolved-sideband regime, $\gamma_m \gg \omega_b$, the vibration mode can still be cooled to its quantum ground state with $N_{st} \simeq 0.5$, in which the conventional sideband-cooling protocols fail. By contrast, without the aid of magnon squeezing a much higher $N_{st}>10^2$ is achieved, as seen in Fig.~\ref{fig2}(i).

\begin{figure}[t]
\hskip-0.9cm\includegraphics[width=0.95\columnwidth]{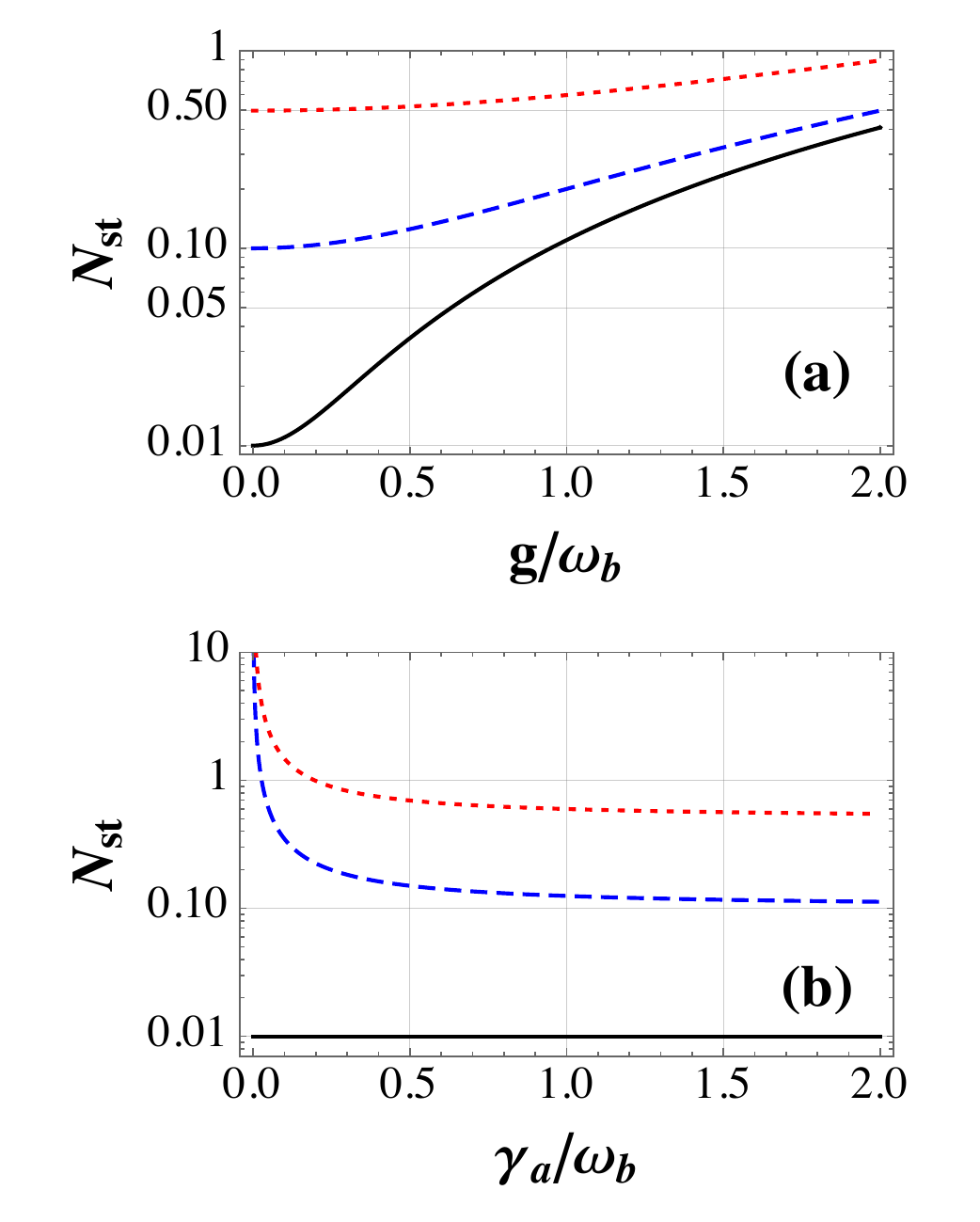}\\
\caption{(a) Steady-state mean phonon number $N_{st}$ versus the normalized coupling $g/\omega_b$ for $\gamma_m=0.1\omega_b$ (black curve), $\gamma_m=\omega_b$ (blue dashed curve), $\gamma_m=5 \omega_b$ (red dotted curve). (b) $N_{st}$ versus the normalized damping of the microwave cavity $\gamma_a/\omega_b$ for $g=0$ (black curve), $g=0.5\omega_b$ (blue dashed curve), $g=\omega_b$ (red dotted curve). We use the optimal detuning $\Delta_a = \omega_b$ and squeezing parameters $|\zeta|^{\rm opt}$ and $\phi^{\rm opt}$. The other parameters are the same as in Fig.~\ref{fig2}.}
\label{fig3}
\end{figure}

Figure~\ref{fig2} is plotted with no coupling to the microwave cavity, i.e., $g=0$. The effects of the magnon-cavity coupling on the cooling performance are studied in Fig.~\ref{fig3}, by plotting $N_{st}$ as a function of the coupling $g$ in Fig.~\ref{fig3}(a) and of the cavity decay rate $\gamma_a$ in Fig.~\ref{fig3}(b). It shows that a vanishing coupling $g \to 0$ and a sufficiently large $\gamma_a \gg \omega_b$ are optimal for achieving the quantum ground state of the vibration mode, which implies that the optimal configuration for realizing mechanical cooling is actually the two-mode magnomechanical system without involving the microwave cavity. This then enables the direct application of optomechanical cooling mechanisms to magnomechanical systems. In this case, the magnon mode is activated by a uniform bias magnetic field and a microwave drive field applied, e.g., via a loop coil~\cite{Jie22,YP}.

Lastly, we discuss how to detect the mechanical cooling. The mechanical state can be measured by coupling the deformation displacement to an additional optical cavity that is driven by a weak red-detuned light~\cite{Jie18,Jie22}. The interaction between light and the mechanical vibration is then a beam-splitter interaction, which maps the mechanical state onto the cavity output field. Thus, by measuring the cavity output field, the final phonon mode occupancy can be detected~\cite{c2}.  It is also beneficial to discuss the feasibility of our proposal especially on the effective squeezing $|\zeta| = 2 \xi |m_s|^2$ that can be achieved in the experiment. For a 250-$\mu$m-diameter YIG sphere, the self-Kerr coefficient $\xi/2\pi \simeq 6.4 \times 10^{-9}$ Hz~\cite{Jie18,YP}, and the magnon excitation number $|m_s|^2$ can be up to $10^{15}$ for drive powers used in \cite{YP}, so the effective squeezing can be up to $|\zeta|/2\pi \simeq 12.8$ MHz. This can satisfy the optimal conditions required in Fig.~\ref{fig2}(c) and \ref{fig2}(f) considering a typical mechanical frequency of $\omega_b/2\pi \simeq 10$ MHz~\cite{Tang,Davis2}. For a significantly large magnon damping $\gamma_m = 5 \omega_b$ used in Fig.~\ref{fig2}(i), one may consider using other materials to have a stronger self-Kerr coefficient and thus a larger squeezing parameter, in order to meet the optimal condition $|\zeta| = \gamma_m$ for mechanical cooling.

\section*{4. Conclusions}

We provide a cooling mechanism for cavity magnomechanical systems by exploiting the squeezing of the magnon mode.  The magnon squeezing can significantly and even completely suppress the magnomechanical Stokes scattering under appropriate conditions, which greatly enhances the cooling efficiency and makes the ground-state cooling of a massive mechanical oscillator possible, even in the unresolved-sideband regime. Although the present work focuses on the self-Kerr induced squeezing, the conclusion that magnon squeezing enhances mechanical cooling is generally valid for other approaches of squeezing, such as magnon squeezing by transferring squeezing from a squeezed-vacuum microwave field~\cite{Jie19a}, or by using the anisotropy of the ferromagnet~\cite{Yuan}.
We also find that the coupling to the microwave cavity hinders the mechanical cooling, and therefore the microwave cavity turns into an unnecessary ingredient in cavity magnomechanics for cooling the vibration mode. The work may find applications in cooling in the dispersively-coupled systems with unresolved sidebands and in the study of macroscopic quantum states.

\section*{Declaration of Competing Interest}

The authors declared that they have no conflict of interest to this work.

\section*{Acknowledgments}

This work has been supported by Zhejiang Province Program for Science and Technology (Grant No. 2020C01019) and the National Natural Science Foundation of China (Grants Nos. U1801661, 11874249, 11934010, 12174329). 

\end{document}